\documentclass{article}
\pdfoutput=1
\usepackage{spconf}
\usepackage{amsmath}
\usepackage{amssymb}
\usepackage{graphicx}
\usepackage{xcolor}
\usepackage{xspace}
\usepackage{amsfonts}
\usepackage{url} % Formatting urls
\usepackage[colorlinks=true]{hyperref} % To follows url links
\usepackage{balance}
\usepackage[ruled]{algorithm2e}
\definecolor{NewTextBG}{rgb}{0.8,0.8,1.00}

\newcommand{\BlackBox}{\rule{1.5ex}{1.5ex}}  % end of proof

\renewcommand{\vec}[1]{\ensuremath{\mathbf{\MakeLowercase{#1}}}}
\newcommand{\mat}[1]{\ensuremath{\mathbf{\MakeUppercase{#1}}}}

\newcommand{\st}{\ensuremath{\quad\mathrm{s.t.}\quad}}
\newcommand{\norm}[1]{\ensuremath{\left\|#1\right\|}}
\newcommand{\cost}[1]{\ensuremath{\ell_{#1}}\xspace}

\newcommand{\setdef}[1]{\ensuremath{\left\{#1\right\}}}

\newcommand{\transp}[1]{^\intercal}
\newcommand{\refeq}[1]{(\ref{#1})}

\def\sgnf{\mathrm{sgn}}

\def\defeq{:=}

\def\data{y}
\def\datam{\mat{\data}}

\def\coef{x}
\def\coefm{\mat{\coef}}

\def\err{e}
\def\errm{\mat{\err}}

\def\rank{k}
\def\ndims{m}
\def\nsamples{n}

\def\matspace{\reals^{\ndims{\times}\nsamples}}
\def\leftspace{\reals^{\ndims{\times}\rank}}
\def\sigmaspace{\reals^{\rank{\times}\rank}}
\def\rightspace{\reals^{\rank{\times}\nsamples}}

\def\reals{\ensuremath{\mathbb{R}}}

\def\rankf{\mathrm{rank}}
\def\diagf{\mathrm{diag}}

\def\modelclass{\mathcal{M}}
\title{\MakeUppercase{Low-rank data modeling via the Minimum Description Length principle}}
\name{Ignacio Ram\'{i}rez and Guillermo Sapiro\thanks{Work supported by NSF, NGA, ONR, DARPA, ARO, and NSSEFF.}}
\address{Department of Electrical and Computer Engineering, University of Minnesota}
\begin{document}
%%%%%%%%%%%%%%%%%%%%%%%%%%%%%%%%%%%%%%%%%%%%%%%%%%%%%%%%%%%%%%%%%%%%%%%%%%%
\ninept
\maketitle
\begin{abstract}
Robust low-rank matrix estimation is a topic of increasing interest, with
promising applications in a variety of fields, from computer vision to data
mining and recommender systems. Recent theoretical results establish the
ability of such data models to recover the true underlying low-rank matrix
when a large portion of the measured matrix is either missing or arbitrarily
corrupted. However, if low rank is not a hypothesis about the true nature of
the data, but a device for extracting regularity from it, no current
guidelines exist for choosing the rank of the estimated matrix. In this work
we address this problem by means of the Minimum Description Length (MDL)
principle -- a well established information-theoretic approach to
statistical inference -- as a guideline for selecting a model for the data
at hand. We demonstrate the practical usefulness of our formal approach with
results for complex background extraction in video sequences.
\end{abstract}

%======================================================================
\section{Introduction}
%======================================================================
\label{sec:introduction}

The key to success in signal processing applications often depends on
incorporating the right prior information about the  data into the
processing algorithms. In matrix estimation, low-rank is an all-time popular
choice, with analysis tools such as Principal Component Analysis (PCA)
dominating the field. However, PCA estimation is known to be non-robust, and
developing robust alternatives is an active research field
(see~\cite{candes11acm} for a review on low-rank matrix estimation).  In
this work, we focus on a recent robust variant of PCA, coined
RPCA~\cite{candes11acm}, which assumes that the difference between the
observed matrix $\datam$, and the true underlying data $\coefm$, is a sparse
matrix $\errm$ whose non-zero entries are arbitrarily valued.  It has been
shown in~\cite{candes11acm} that $\coefm$ (alternatively, $\errm$) can be
recovered exactly by means of a convex optimization problem involving the
rank of $\datam$ and the \cost{1} norm of $\errm$. The power of this
approach has been recently demonstrated in a variety of applications, mainly
computer vision (see~\cite{wright09nips} and
{\scriptsize\url{http://perception.csl.uiuc.edu/matrix-rank/applications.html}}
for examples).

However, when used as a pure data modeling tool, with no assumed ``true''
underlying signal, the rank of $\coefm$ in a PCA/RPCA decomposition is a
parameter to be tuned in order to achieve some desired goal. A typical case
is \emph{model selection}~\cite[Chapter~7]{hastie09}, where one wants to
select the \emph{size} of the model (in this case, rank of the
approximation) in order to strike an optimal balance between the ability of
the estimated model to generalize to new samples, and its ability to adapt
itself to the observed data (the classic overfitting/underfitting trade-off
in statistics). The main issue in model selection is how to formulate this
balance as a cost function.

In this work, we address this issue via the Minimum Description Length (MDL)
principle~\cite{rissanen78,barron98}.\footnote{While here we address the
  matrix formulation, the developed framework is applicable in general,
  including to sparse models, and such general formulation will be reported
  in our extended version of this work.} MDL is a general methodology for
assessing the ability of statistical models to capture regularity from
data. The MDL principle can be regarded as a practical implementation of the
Occam's razor principle, which states that, given two descriptions for a
given phenomenon, the shorter one is usually the best. In a nutshell, MDL
equates ``ability to capture regularity'' with ``ability to compress'' the
data, using \emph{codelength} or \emph{compressibility} as the metric for
measuring candidate models.

The resulting framework provides a robust, parameter-free low-rank matrix
selection algorithm, capable of capturing relevant low-rank information in
the data, as in the video sequences from surveillance cameras in the
illustrative application here reported. From a theoretical standpoint, this
brings a new, information theoretical perspective into the problem of
low-rank matrix completion. Another important feature of an MDL-based
framework such as the one here presented is that new prior information can
be naturally and easily incorporated into the problem, and its effect can be
assessed \emph{objectively} in terms of the different codelengths obtained.

%======================================================================
\section{Low-rank matrix estimation/approximation}
%======================================================================
\label{sec:lr-intro}

Under the low-rank assumption, a matrix $\datam \in \matspace$ can be
written as $\datam=\coefm+\errm$, where $\rankf(\coefm) \ll
\min\{\ndims,\nsamples\}$ and $\norm{\errm} \ll \norm{\datam}$, where
$\norm{\cdot}$ is some matrix norm. Classic PCA provides the best
rank-$\rank$ approximation to $\datam$ under the assumption that $\errm$ is
a random matrix with zero-mean IID Gaussian entries,
\begin{equation}
\hat\coefm = \arg\min_\mat{W} \norm{\datam-\mat{W}}_2,\,\st\,\rankf(\mat{W}) \leq \rank.
\label{eq:pca}
\end{equation}
However, PCA is known to be non-robust, meaning that the estimate
$\hat\coefm$ can vary significantly if only a few coefficients in $\errm$
are modified. This work, providing an example of introducing the MDL
framework in this type of problems, focuses on a robust variant of PCA,
RPCA, introduced in~\cite{candes11acm}. RPCA estimates $\coefm$ via the
following convex optimization problem,
\begin{equation}
\hat\coefm = \arg\min_\mat{W} \norm{\datam-\mat{W}}_1 + \lambda\norm{\mat{W}}_{*},
\label{eq:rpca}
\end{equation}
where $\norm{\mat{W}}_{*} \defeq \sum_i \sigma(\mat{W})_i$ is the nuclear
norm of $\mat{W}$ ($\sigma(\mat{W})_i$ denotes the $i$-th singular value of
$\mat{W}$). The rationale behind \refeq{eq:rpca} is as follows. First, the
\cost{1} fitting term allows for large errors to occur in the
approximation. In this sense, it is a robust alternative to the \cost{2}
norm used in PCA.  The second term, $\lambda\norm{\mat{W}}_{*},$ is a convex
approximation to the PCA constraint $\rankf(\mat{W}) \leq \rank$, merged
into the cost function via a Lagrange multiplier $\lambda$.

This formulation has been recently shown to be notoriously robust, in the
sense that, if a true low-rank matrix $\coefm$ exists, it can be recovered
using \refeq{eq:rpca} even when a significant amount of coefficients in
$\errm$ are arbitrarily large~\cite{candes11acm}. This can be achieved by
setting $\lambda=1/\sqrt{\max \{\ndims,\nsamples\}}$, so that the procedure
is parameter-free.

\subsection{Low-rank approximation as dimensionality reduction}

In many applications, the goal of low-rank approximation is not to find a
``true'' underlying matrix $\coefm$, but to perform what is known as
``dimensionality reduction,'' that is, to obtain a succinct representation
of $\datam$ in a lower dimensional subspace. A typical example is feature
selection for classification. In such cases, $\errm$ is not necessarily a
small measurement perturbation, but a \emph{systematic}, possibly large,
error derived from the approximation process itself. Thus, RPCA arises as an
appealing alternative for low-rank approximation.

However, in the absence of a true underlying signal $\coefm$ (and deviation
$\errm$), it is not clear how to choose a value of $\lambda$ that produces a
good approximation of the given data $\datam$ for a given application. A
typical approach would involve some cross-validation step to select
$\lambda$ to maximize the final results of the application (for example,
minimize the error rate in a classification problem).

The issue with cross-validation in this situation is that the best model is
selected \emph{indirectly} in terms of the final results, which can depend
in unexpected ways on later stages in the data processing chain of the
application (for example, on some post-processing of the extracted
features). Instead, we propose to select the best low-rank approximation by
means of a \emph{direct measure} on the intrinsic ability of the resulting
model to capture the desired regularity from the data, this also providing a
better understanding of the actual structure of the data. To this end, we
use the MDL principle, a general information-theoretic framework for model
selection which provides means to define such a direct measure.
%
%======================================================================
\section{MDL-based low-rank model selection}
%======================================================================
\label{sec:mdl-lr}

Consider a family $\modelclass$ of candidate models which can be used to
describe a matrix $\datam$ \emph{exactly} (that is, losslessly) using some
encoding procedure. Denote by $L(\datam|M)$ the description length, in bits,
of $\datam$ under the description provided by a given model $M \in
\modelclass$.  MDL will then select the model $\hat{M} \in \modelclass$ for
$\datam$ for which $L(\datam|\hat{M})$ is minimal, that is
$\hat{M}=\arg\min_{M \in \modelclass} L(\datam|M)$.  It is a standard
practice in MDL to use the \emph{ideal} Shannon code for translating
probabilities into codelengths. Under this scheme, a sample value $\data$
with probability $P(\data)$ is assigned a code with length $L(\data) = -\log
P(\data)$ (all logarithms are taken on base $2$). This is called an ideal
code because it only specifies a codelength, not a specific binary code, and
because the codelengths produced can be fractional.

By means of the Shannon code assignment, encoding schemes $L(\cdot)$ can be
defined naturally in terms of probability models $P(\cdot)$. Therefore, the
art of applying MDL lies in defining appropriate probability assignments
$P(\cdot)$, that exploit as much prior information as possible about the
data at hand, in order to maximize compressibility.  In our case, there are
two main components to exploit. One is the low-rank nature of the
approximation $\coefm$, and the other is that most of the entries in $\errm$
will be small, or even zero (in which case $\errm$ will be \emph{sparse}).
Given a low-rank approximation $\coefm$ of $\datam$, we describe $\datam$ as
the pair $(\coefm,\errm)$, with $\errm=\datam-\coefm$. Thus, our family of
models is given by $\modelclass=\setdef{(\coefm,\errm): \datam=\coefm+\errm,
  \rankf(\coefm) \leq \rankf(\datam)}$.  As $\errm=\datam-\coefm$, we index
$\modelclass$ solely by $\coefm$. With these definitions, the description
codelength of $\datam$ is given by $L(\datam|\coefm) \defeq L(\coefm) +
L(\errm)$.  Now, to exploit the low rank of $\coefm$, we describe
it in terms of its reduced SVD decomposition,
\begin{equation}
\coefm=\mat{U}\Sigma\mat{V}\transp ,\;\mat{U} \in \leftspace,\; \Sigma \in \sigmaspace,\; \mat{V} \in \rightspace,
\end{equation}
where $\rank$ is the rank of $\coefm$ (the zero-eigenvalues and the
respective left and right eigenvectors are discarded in this
description). We now have $L(\coefm)=L(\mat{U}) + L(\Sigma) +
L(\mat{V})$. Clearly, such description will be short if $\rankf(\coefm)$ is
significantly smaller than $\max\{\ndims,\nsamples\}$. We may also be able
to exploit further structure in $\mat{U}$, $\Sigma$ and $\mat{V}$.  

\subsection{Encoding $\Sigma$}

The diagonal of $\Sigma$ is a non-increasing sequence of
$\rank$ positive values. However, no safe assumption can be made about the
magnitude of such values. For this scenario we propose to use the
\emph{universal prior for integers}, a general scheme for encoding arbitrary
positive integers in an efficient way~\cite{rissanen92},
\begin{equation}
L(j) = \log^* j \defeq \log j + \log \log j + \ldots + \log 2.865,
\label{eq:universal-integers}
\end{equation}
where the sum stops at the first non-positive summand, and $\log 2.865$ is
added to satisfy Kraft's inequality (a requirement for the code to be
uniquely decodable, see~\cite[Chapter~5]{cover06}). In order to apply
\refeq{eq:universal-integers}, the diagonal of $\Sigma$, $\diagf(\Sigma)$, is
mapped to an integer sequence via $[10^{16}\diagf(\Sigma)]$, where $[\cdot]$
denotes rounding to nearest integer (this is equivalent to quantizing
$\diagf(\Sigma)$ with precision $\delta_\Sigma=10^{-16}$).

\subsection{Encoding {\bf U} and {\bf V}, general case}

By virtue of the SVD algorithm, the columns of $\mat{U}$ and $\mat{V}$ have
unit norm.  Therefore, the most general assumption we can make about
$\mat{U}$ and $\mat{V}$ is that their columns lie on the respective
$\ndims$-dimensional and $\nsamples$-dimensional unit spheres.  

\begin{figure}
\begin{center}
\includegraphics[height=1.4in]{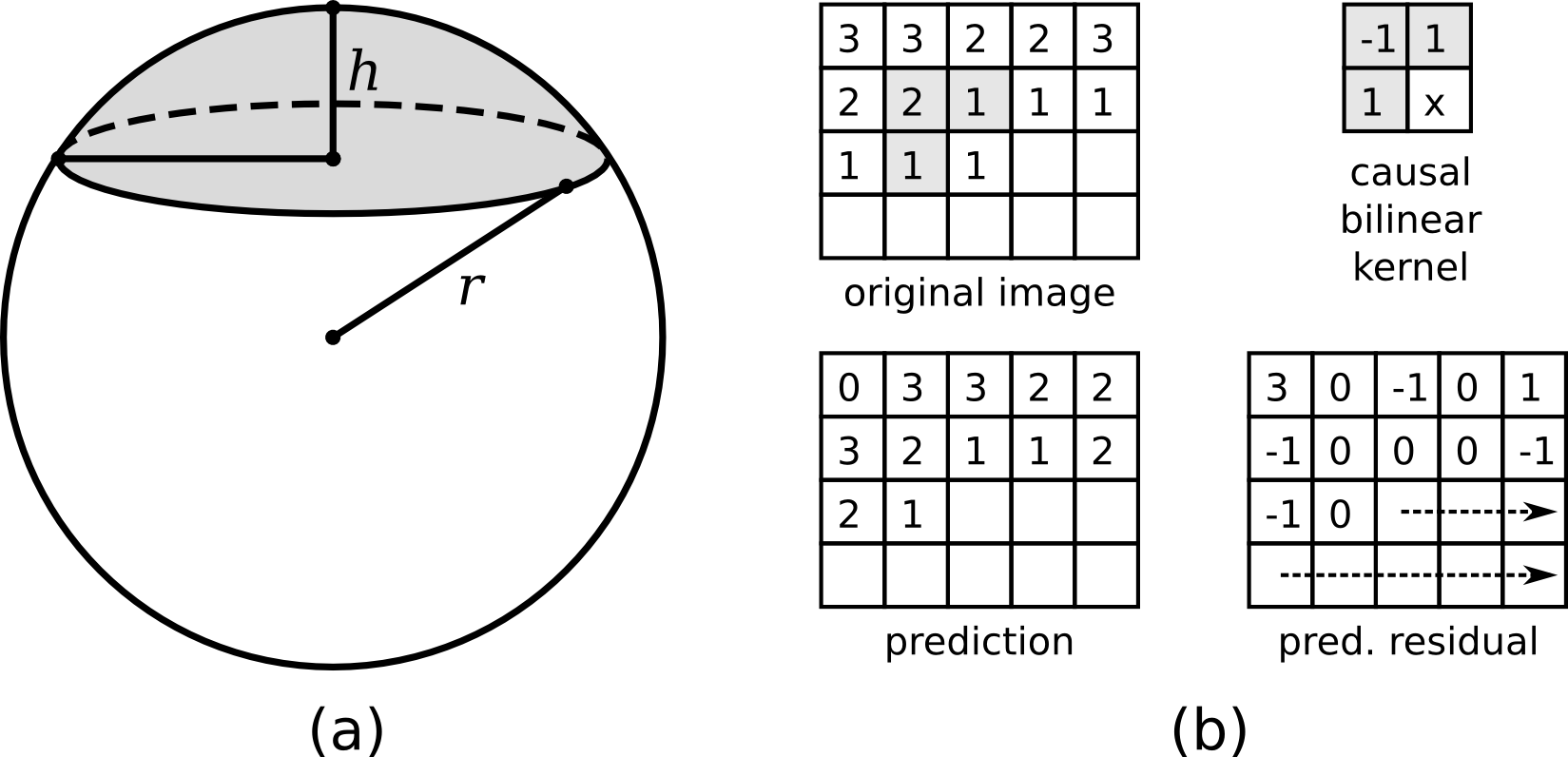}%
\end{center}\vspace{-4ex}
\caption{\label{fig:encoding-scheme} (a) The spherical cap of radius
  $\mathrm{r}$ and  height $\mathrm{h}$ (shown in
  gray). (b) Causal bilinear prediction of smooth 2D images. }
\end{figure}

An efficient code for this case can be obtained by encoding each column of
$\mat{U}$ and $\mat{V}$ in the following manner. Let $\vec{u}_i$ be a column
of $\mat{U}$ ($\mat{V}$ is similarly encoded). Since $\vec{u}_i$ is assumed
to be distributed uniformly over the $\ndims$-dimensional unit sphere, the
marginal cumulative density function of the first element $u_{1i}$,
$F(u_{1i})=P(x \leq u_{1i})$, corresponds to the proportion of vectors
$\vec{u}$ that lie on the \emph{unit spherical cap} of height $h=1+u_{1i}$
(see Figure~\ref{fig:encoding-scheme}(a)). This proportion is given by
 $F(u_{1i}) =A_\ndims(1+u_{1i},1)/S_\ndims(1)$ where $A_\ndims(h,r)$
and $S_\ndims(r)$ are the area of spherical cap of height $h$ and the total
surface area of the $\ndims$-dimensional sphere of radius $r$ respectively.
These are given for the case $0 \leq h \leq r$ ($-1 \leq u_{1i} \leq 0$) by
(see~\cite{li2011ajms}),
\begin{eqnarray*}
A_\ndims(h,r)&=&\frac{1}{2}S_\ndims(r)I((2hr-h^2)/r^2\,;\,\frac{\ndims-1}{2},\frac{1}{2})\\
S_\ndims(r) &=& 2\pi^{\ndims/2}r^{\ndims-1}\Gamma^{-1}(\ndims/2),
\end{eqnarray*}
where $I(x\,;\,a,b)=\frac{\int_{0}^{x}t^{a-1}(1-t)^{b-1}dt}{B(a,b)}$, and
$B(a,b)=\int_{0}^{1}t^{a-1}(1-t)^{b-1}dt$ are the regularized incomplete
Beta function and the Beta function of parameters $a,b$ respectively, and
$\Gamma(\cdot)$ is the Gamma function.  When $r < h \leq 2r$ we simply have
$A_\ndims(h,r)=1-A_\ndims(2r-h,r)$. For encoding $u_{1i}$ we have $r=1$ so that
\begin{equation}
F(u_{1i})  = (1/2)I(1-u_{1i}^2;(\ndims-1)/2,1/2), -1 \leq u_{1i} \leq 0,
\end{equation}
since $2h-h^2=h(2-h)=(1+u_{1i})[2-(1+u_{1i})]=1-u_{1i}^2$. Finally, we
compute the Shannon codelength for $u_{1i}$ as
\begin{align*}
p(u_{1i}) &= F'(u_{1i}) \stackrel{(a)}{=} \frac{(1-u_{1i}^2)^{(\ndims-3)/2}(u_{1i}^2)^{-1/2}}{2\cdot B\left(\frac{\ndims-1}{2},\frac{1}{2}\right) }(-2u_{1i}) \\
&= -\sgnf(u_{1i}) (1-u_{1i}^2)^{(\ndims-3)/2}[B((\ndims-1)/2,1/2)]^{-1} \\
-\log p(u_{1i}) &=
 -\frac{\ndims-3}{2}\log (1-u_{1i}^2) + \log B((\ndims-1)/2,1/2),
\end{align*}
where in $(a)$ we applied the Fundamental Theorem of Calculus to the
definition of $F(h)$ and the chain rule for derivatives.

With $u_{1i}$ encoded, the vector $(u_{2i},u_{3i},\ldots,u_{\ndims i})$ is
uniformly distributed on the surface of the $(\ndims-1)$-dimensional sphere
of radius $r'=1-|u_{1i}|$, and we can apply the same formula to compute the
probability of $u_{2i}$, $F(u_{2i}) =
A_{\ndims-1}(u_{2i}+r',r')/S_{\ndims-1}(r')$.

Finally, to encode the next column $\vec{u}_{i+1}$, we can exploit its
orthogonality with respect to the previous ones and encode it as a vector
distributed uniformly over the $\ndims-i$ dimensional sphere corresponding
to the intersection of the unit sphere and the subspace perpendicular to $[
  \vec{u}_1,\ldots,\vec{u}_{i}]$.

In order to produce finite descriptions $L(\mat{U})$ and $L(\mat{V})$, both
$\mat{U}$ and $\mat{V}$ also need to be quantized. We choose the
quantization steps for $\mat{U}$ and $\mat{V}$ adaptively, using as a
starting point the empirical standard deviation of a normalized vector, that
is, $\delta_u=\sqrt{1/\ndims}$ and $\delta_v=\sqrt{1/\nsamples}$
respectively, and halving these values until no further decrease in the
overall codelength $L(\datam|\coefm)$ is observed.

\subsection{Encoding {\bf U} predictively}

If more prior information about $\mat{U}$ and $\mat{V}$ is available, it
should be used as well. For example, in the case of our example application,
the columns of $\datam$ are consecutive frames of a video surveillance
camera. In this case, the columns of $\mat{U}$ represent ``eigen-frames'' of
the video sequence, while $\mat{V}$ contains information about the evolution
in time of those frames (this is clearly observed in figures~\ref{fig:lobby}
and~\ref{fig:shopping}). Therefore, the columns of $\mat{U}$ can be assumed
to be piecewise smooth, just as normal static images are. To exploit this
smoothness, we apply a predictive coding to the columns of
$\mat{U}$. Concretely, to encode the $i$-th column $\vec{u}_i$ of $\mat{U}$,
we reshape it as an image $\mat{B}$ of the same size as the original frames
in $\datam$. We then apply a causal bilinear predictor to produce an
estimate of $\mat{B}$, $\mat{\hat B}=\{\hat{b}_{jl}\}$ where
$\hat{b}_{jl}=b_{jl}-b_{j(l-1)}-b_{(j-1)l} + b_{(j-1)(l-1)}$, assuming
out-of-range pixels to be $0$. The prediction residual
$\mat{\tilde{B}}=\mat{B}-\mat{\hat B}$ is then encoded in raster scan as
a sequence of Laplacian random variables with unknown parameter $\theta_u^i$.
This encoding procedure, common in predictive coding, is depicted in
Figure~\ref{fig:encoding-scheme}(b).

Since the parameters $\setdef{\theta_u^i,i=1,\ldots,\rank}$ are unknown, we
need to encode them as well to produce a complete description of
$\datam$. In MDL, this is done using the so-called \emph{universal encoding
  schemes}, which can be regarded as a generalization of classical Shannon
encoding to the case of distributions with unknown parameters
(see~\cite{barron98} for a review on the subject). In this work we adopt the
so-called \emph{universal two-part codes}, and apply it to encode each
column $\vec{u}_i$ separately. Under this scheme, the unknown Laplacian
parameter for $\theta_u^i$ is estimated via Maximum Likelihood,
$\hat\theta_u^i(\vec{u}_i)$, and quantized with precision $1/\sqrt{\ndims}$,
thus requiring $L(\hat\theta_u^i)=\frac{1}{2}\log \ndims + c_1$
bits. Given the quantized $\hat\theta_u^i$, $\vec{u}_i$ is described using
the discretized Laplacian distribution $L(\vec{u}_i) = -\log
P(\vec{u}_i|\hat\theta_u^i(\vec{u}_i)) + c_2$. Here $c_1$ and $c_2$ are
constants which can be disregarded for optimization purposes. It was shown
in~\cite{rissanen78} that the precision $1/\sqrt{\ndims}$ asymptotically yields
the shortest two-parts codelength.

\subsection{Encoding {\bf V} predictively}

We also expect a significant redundancy in the time dimension, so that the
columns of $\mat{V}$ are also smooth functions of time (in this case, sample
index $j=1,2,\ldots,\nsamples$). In this case, we apply a first order causal
predictive model to the columns of $\mat{V}$, by encoding them as sequences
of prediction residuals, $\tilde{\mathbf{v}}_{i} =
(\tilde{v}_{i1},\tilde{v}_{i2},\ldots,\tilde{v}_{i\nsamples})$, with
$\tilde{v}_{ij}={v}_{ij}-{v}_{i(j-1)}$ for $j > 1$ and
$\tilde{v}_{i1}={v}_{i1}$. Each predicted column $\vec{v}_i$ is encoded as a
sequence of Laplacian random variables with unknown parameter
$\theta_v^i$. As with $\mat{U}$, we use a two-parts code here to describe
the data and the unknown Laplacian parameters together. This time, since the
length of the columns is $\nsamples$, the codelength associated to each
$\theta_v^i$ is $L(\theta_v^i) = \frac{1}{2}\log \nsamples$.

\subsection{Encoding {\bf E}}
We exploit the (potential) sparsity of $\errm$ by first describing the
indexes of its non-zero locations using an efficient universal two-parts
code for Bernoulli sequences known as Enumerative Code~\cite{cover73}, and
then the non-zero values at those locations using a Laplacian model. In the
specific case of the experiments of Section~\ref{sec:results}, we encode
each row of $\errm$ separately. Because each row of $\errm$ corresponds to
the pixel values at a fixed location across different frames, we expect some
of these locations to be better predicted than others (for example,
locations which are not occluded by people during the sequences), so that
the variance of the error (hence the Laplacian parameter) will vary
significantly from row to row. As before, the unknown parameters here are
dealt with using a two-parts coding scheme.

\subsection{Model selection algorithm}
\label{sec:algorithm}

To obtain the family of models $\modelclass$ corresponding to all possible
low-rank approximations of $\datam$, we apply the RPCA
decomposition~\refeq{eq:rpca} for a decreasing sequence of values of
$\lambda$, $\{\lambda_t:t=1,2,\ldots\}$ obtaining a corresponding
sequence of decompositions $\{(\coefm_t,\errm_t), t=1,2,\ldots\}$. We obtain
such sequence efficiently by solving \refeq{eq:rpca} via a simple
modification of the Augmented Lagrangian-based (ALM) algorithm proposed
in~\cite{lin09arxiv} to allow for \emph{warm restarts}, that is, where the
initial ALM iterate for computing $(\coefm_t,\errm_t)$ is
$(\coefm_{t-1},\errm_{t-1})$. We then select the pair $(\coefm_{\hat
  t},\errm_{\hat t})$, $\hat{t} = \arg\min_t \{L(\coefm_t)+L(\errm_t)\}$.

%======================================================================
\section{Results and conclusion}
%======================================================================
\label{sec:results}

In order to have a reference base, we repeated the experiments performed in
\cite{wright09nips} using our algorithm. These experiments consist of frames from
surveillance cameras which look at a fixed point where people pass by.
The idea is that, if frames are stacked as columns of $\datam$, the
background can be well modeled as a low-rank component  of $\datam$ ($\coefm$), while
the people passing by appear as ``spurious errors'' ($\errm$). Clearly, if the
background in all frames is the same, it can be very well modeled as a
rank-$1$ matrix where all the columns are equal. However, lighting changes,
shadows, and reflections, ``raise'' the rank of the background, and the
appropriate rank needed to model the background is no longer obvious.

\begin{figure}[t]
\begin{center}
\includegraphics[width=0.95\columnwidth,height=3.4in]{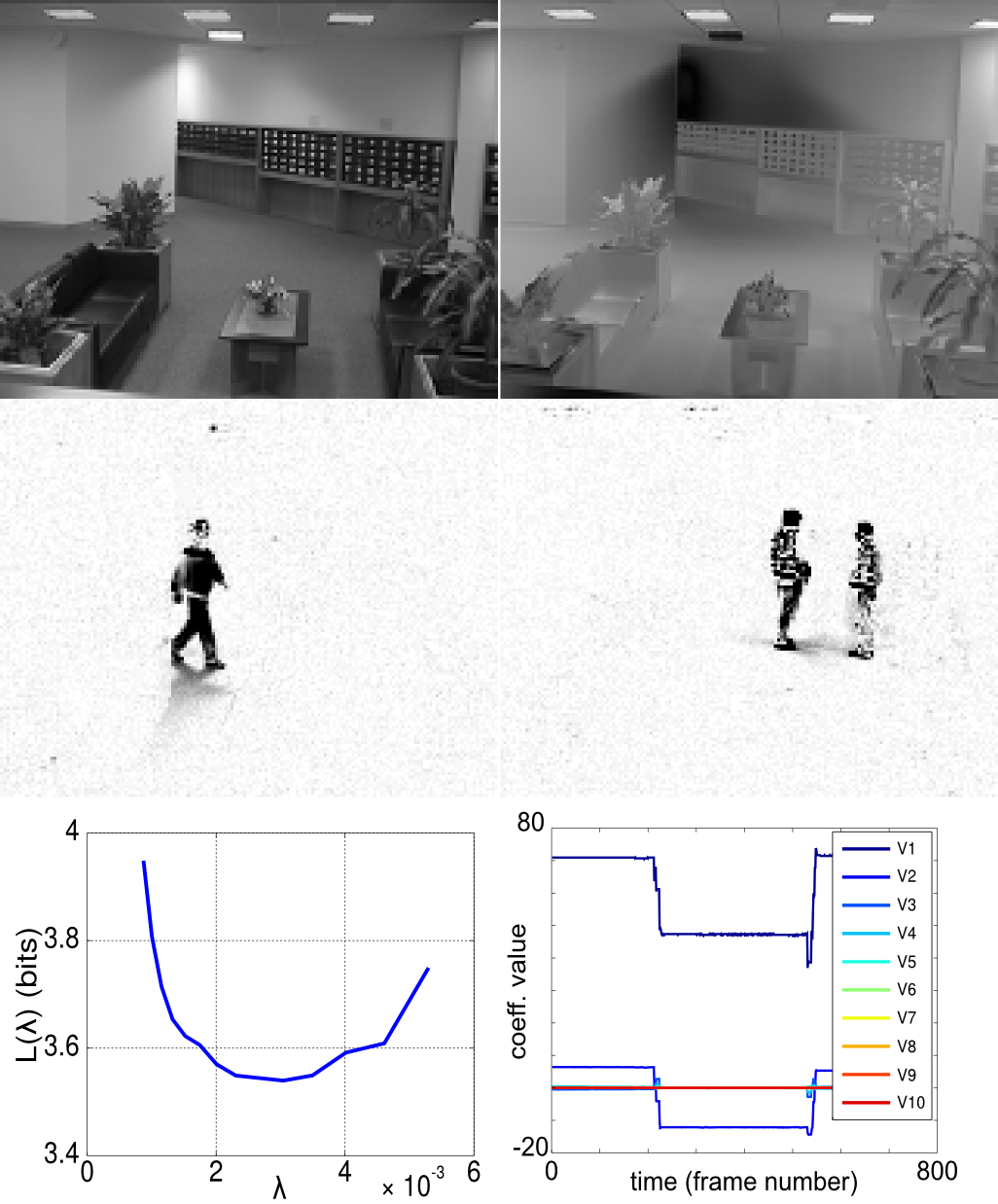}\vspace{-2ex}
\caption{\label{fig:lobby}Results for the ``Lobby'' sequence (see text for a
  description of the above pictures and graphs). The rank of the
  approximation decomposition for this case is $\rank=10$.  The moment where
  the lights are turned off is clearly seen here as the ``square pulse'' in
  the middle of the first two right-eigenvectors (bottom-right figure).  Also
  note how $\vec{u}_2$ (top-right) compensates for changes in shadows.}
\end{center}
\end{figure}

\begin{figure}[t]
\begin{center}
\includegraphics[width=0.95\columnwidth,height=3.4in]{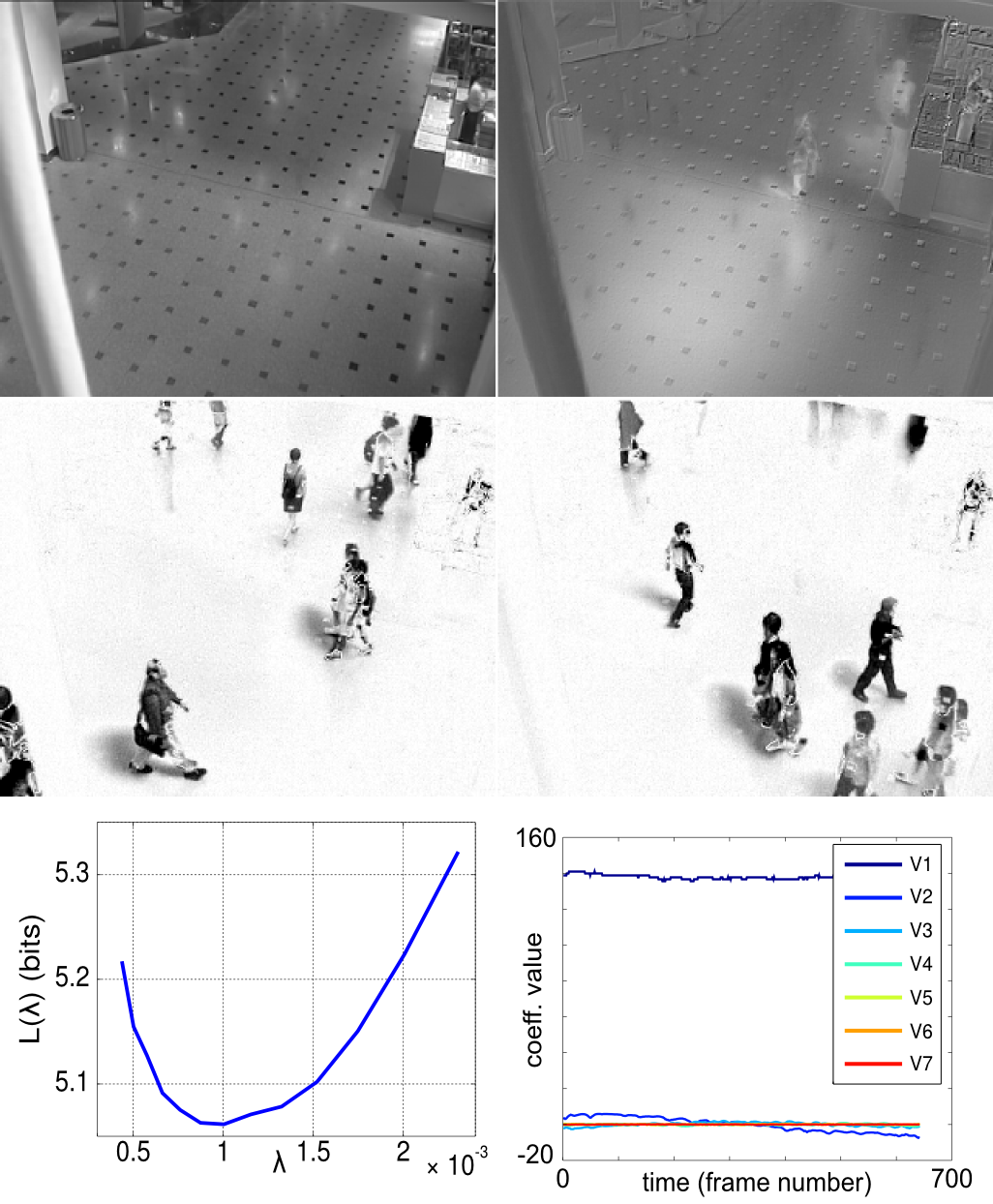}\vspace{-2ex}
\caption{\label{fig:shopping}Results for the ``ShoppingMall'' sequence (see
  text for a description of the above pictures and graphs). In this case,
  the rank of the approximation decomposition is $\rank=7$. Here, the first
  left-eigenvector models the background, whereas the rest tend to capture
  people that stood still for a while. Here we see the ``phantom'' of two
  such persons in the second left-eigenvector (top-right).}
\end{center}
\end{figure}

Concretely, the experiments here described correspond to two sequences:
``Lobby'' and ``ShoppingMall,'' whose corresponding results are summarized
respectively in figures~\ref{fig:lobby} and~\ref{fig:shopping}.\footnote{The
  full videos can be viewed at \url{http://www.tc.umn.edu/~nacho/lowrank/}.}
At the top of both figures, the first two left-eigenvectors $\vec{u}_1$ and
$\vec{u}_2$ of $\coefm$ are shown as 2D images. The middle shows two sample
frames of the error approximation.  The $L$-vs-$\lambda$ curve is shown at
the bottom-left (note that the best $\lambda$ is \emph{not} the one dictated
by the theory in~\cite{candes11acm}, which are $0.007$ for Lobby and
$0.0035$ for ShoppingMall, both outside of the plotted range), and the
scaled right-eigenvectors $\sigma_i \vec{v}_i$ are shown on the
bottom-right. In both cases, the resulting decomposition recovered the
low-rank structure correctly, including the background, its changes in
illumination, and the effect of shadows. It can be appreciated in the
figures~\ref{fig:lobby}-\ref{fig:shopping} how such approximations are
naturally obtained as combinations of a few significant eigen-vectors,
starting with the average background, followed by other details.

\subsection{Conclusion}
In summary, we have presented an MDL-based framework for low-rank data
approximation, which combines state-of-the-art algorithms for robust
low-rank decomposition with tools from information theory.  This framework
is able to capture the underlying low-rank information on the experiments
that we performed, out of the box, and without any hand parameter tuning,
thus constituting a promising competitive alternative for automatic data
analysis and feature extraction.
%
%======================================================================
%\include{icassp2012}
\bibliography{icassp2012}
\bibliographystyle{IEEEbib}
%======================================================================

%%%%%%%%%%%%%%%%%%%%%%%%%%%%%%%%%%%%%%%%%%%%%%%%%%%%%%%%%%%%%%%%%%%%%%%%%%%
\end{document}